\documentclass[a4paper]{article}
\usepackage[pdftex]{graphicx}
\usepackage{amsmath,amssymb,amsthm}

\usepackage{triotex}
\usepackage{listings}
\usepackage{xcolor}
\usepackage[pdftex]{hyperref}

\usepackage{times}

\usepackage{tikz}
\usetikzlibrary{arrows,automata,shapes,shapes.symbols,positioning}

\theoremstyle{plain}
\newtheorem{theorem}{Theorem}

\theoremstyle{definition}
\newtheorem{definition}[theorem]{Definition}

\newcommand{\lmk}{\mathop{\rhd}}
\newcommand{\rmk}{\mathop{\lhd}}
\newcommand{\emk}{\rmk}

\newcommand{\mk}{\downarrow}
\newcommand{\mkat}[1]{\downarrow\!\!(#1)}
\newcommand{\powerset}[1]{\wp(#1)}

\newcommand{\nfa}[1]{\TRIOitem{NFA}{#1}}
\newcommand{\dfa}[1]{\TRIOitem{DFA}{#1}}
\newcommand{\nfal}[1]{\nfa{#1}}
\newcommand{\dfal}[1]{\dfa{#1}}

\newcommand{\dt}[1]{\mathcal{D}_{#1}}
\newcommand{\ndt}[1]{\mathcal{N}_{#1}}
\newcommand{\rdt}[1]{\mathcal{R\!D}_{#1}}
\newcommand{\rndt}[1]{\mathcal{R\!N}_{#1}}

\newcommand{\lang}[1]{\mathcal{L}(#1)}

\newcommand{\Sigmaext}{\Sigma_{\mathsf{\lmk\!\rmk}}}
\newcommand{\Sigmaend}{\Sigma_{\emk}}

\newcommand{\pad}{\Box}
\newcommand{\ift}[1]{\textsc{#1}}
\newcommand{\rb}[1]{\stackrel{\curvearrowleft}{#1}}

\newcommand{\inact}{\mathcal{I}}
\newcommand{\act}{\mathcal{H}}

\newcommand{\proj}[2]{{\exists_#2}{#1}}
\newcommand{\fall}[2]{{\forall_#2}{#1}}

\newcommand{\cat}{\mathbin{\circ}}
\newcommand{\rev}[1]{\mathit{rev}(#1)}
\newcommand{\comp}[1]{\overline{#1}}
\newcommand{\conv}{\mathbin{\otimes}}
\newcommand{\convof}[1]{{#1}^{\mathop{\otimes}}}

\begin{document}

\title{A Survey of Multi-Tape Automata}

\author{Carlo A. Furia}

\date{May 2012}


\maketitle

\vspace{1cm}
\begin{abstract}
This paper summarizes the fundamental expressiveness, closure, and decidability properties of various finite-state automata classes with multiple input tapes.
It also includes an original algorithm for the intersection of one-way nondeterministic finite-state automata.
\end{abstract}

\newpage 

\tableofcontents

\newpage

\section{Overview}
This paper is a survey of the known results about the expressiveness, closure, and decidability properties of finite-state automata that read multiple input tapes.
The theoretical study of these computing devices began within the classic work of Rabin and Scott~\cite{RabinScott59}, Elgot and Mezei~\cite{ElgotMezei65}, and Rosenberg~\cite{Rosenberg-PhD,Rosenberg67,FischerR68}.
In recent years, multi-tape automata have found some applications such as automatic structures~\cite{KhoussainovN94,Shapiro92,BlumensathG00,Rubin08}, querying string databases~\cite{GrahneNU99}, and weighted automata for computational linguistic~\cite{KGN04,ChamparnaudGKN08}.

Section~\ref{sec:definitions} gives a general definition of multi-tape automata that allows for nondeterministic, asynchronous, and two-way movements of the input heads.
The following sections analyze the expressiveness, closure properties, and decidability for increasing levels of generality: Section~\ref{sec:single-tape-automata} briefly recalls the well-known properties of single-tape automata; Section~\ref{sec:synchr-multi-tape} discusses synchronous automata, where different input heads are not allowed to be in arbitrary portions of their respective tapes; and Section~\ref{sec:asynchr-multi-tape} considers the most general case of fully asynchronous heads.
Section~\ref{sec:summary-properties} gives a synopsis of the fundamental properties surveyed.

Finally, Section~\ref{sec:intersection} presents an original algorithm for the intersection of asynchronous one-way multi-tape automata.
Since this class is not closed under intersection, the algorithm may not terminate in the general case.

\section{Definitions} \label{sec:definitions}

\subsection{Alphabets, Words, and Tuples}
$\integers$ is the set of integer numbers, and $\naturals$ is the set of natural numbers $0, 1, \ldots$.
For a (finite) set $S$, $\powerset{S}$ denotes its powerset.
For a finite nonempty \emph{alphabet} $\Sigma$, $\Sigma^*$ denotes the set of all finite sequences $\sigma_1\, \ldots\, \sigma_n$, with $n \geq 0$, of symbols from $\Sigma$ (also called \emph{words} over $\Sigma$); when $n = 0$, $\epsilon \in \Sigma^*$ is the \emph{empty} word.
$|s| \in \naturals$ denotes the length $n$ of a word $s = \sigma_1 \,\ldots\, \sigma_n$.
An \emph{$n$-word} is an $n$-tuple $\langle x_1, \ldots, x_n \rangle \in (\Sigma^*)^n$ of words over $\Sigma$, where each $x_i = \sigma_{1,i}\,\ldots\,\sigma_{n_i,i}$.

The \emph{convolution} of $n$ words $\langle x_1, \ldots, x_n \rangle$ is a word over the padded alphabet $(\Sigma \cup \{\pad\})^n$:
\begin{equation*}
x_1 \conv \cdots \conv x_n = 
\left[
\begin{array}{c}
y_1^1 \\
\vdots \\
y_n^1
\end{array}
\right]
\cdots
\left[
\begin{array}{c}
y_1^m \\
\vdots \\
y_n^m
\end{array}
\right]\,,
\end{equation*}
where $m$ is the maximum length $\max \{|x_1|, \ldots, |x_n|\}$, and 
\begin{equation*}
y_k^h  \ =\ 
\begin{cases}
\text{the } h\text{-th component of } x_{h,k} \text{ of } x_k  &  h \leq |x_k|\,,  \\
\pad & \text{otherwise}\,.
\end{cases}
\end{equation*}

Given two words $x = x_1\, \ldots\, x_n$ and $y = y_1\, \ldots\, y_m$, $x \cat y$ denotes their \emph{concatenation} $x_1\, \ldots\,x_n\,y_1\,\ldots\,y_m$, and $\rev{x}$ denotes the \emph{reversal} $x_n\,\ldots\,x_1$ of $x$.
These operations on words are naturally lifted to sets of words and to $n$-words; for example, $W \cat X = \{w\cat x \mid w \in W, x \in X\}$ denotes the concatenation of the two sets of words $W$ and $X$; $\rev(x) = \langle \rev{x_1}, \ldots, \rev{x_n}\rangle$ denotes the reversal of the $n$-word $x = \langle x_1, \ldots, x_n \rangle$.
Also, $W^*$ denotes the \emph{Kleene closure} with respect to finite self-concatenation, that is $W^* = \bigcup_{k\in\naturals}W^k$, where 
$$
W^k \ =\ \overbrace{W \cat W \cat \cdots \cat W}^{k} \,.
$$

Given a set $W$ of $n$-words over $\Sigma$, $\comp{W}$ is the \emph{complement} set $(\Sigma^*)^n \setminus W$; $\proj{W}{k}$ is the \emph{projection} set of $n-1$-words obtained by projecting out $W$'s $k$-th component, that is 
$$
\proj{W}{k} \ =\ 
\left\{
\langle y_1, \ldots, y_{n-1} \rangle \mid 
\begin{array}{l}
\exists y \langle x_1, \ldots, x_{k-1}, y, x_{k+1}, \ldots x_n \rangle \in W, \\
x_1 = y_1, \ldots, x_{k-1} = y_{k-1}, \\
x_{k+1} = y_k, \ldots, x_{n} = y_{n-1}
\end{array}
\right\}
 \,;
$$
$\fall{W}{k}$ is the \emph{generalization} set of $n-1$-words obtained for every value of $W$'s $k$-th component, that is 
$$
\fall{W}{k} \ =\ 
\left\{
\langle y_1, \ldots, y_{n-1} \rangle \mid 
\begin{array}{l}
\forall y \langle x_1, \ldots, x_{k-1}, y, x_{k+1}, \ldots x_n \rangle \in W, \\
x_1 = y_1, \ldots, x_{k-1} = y_{k-1}, \\
x_{k+1} = y_k, \ldots, x_{n} = y_{n-1}
\end{array}
\right\}
 \,;
$$
and the convolution of $W$ is the set of words
$$
\convof{W} \ =\ \{ x_1 \conv \cdots \conv x_n \mid \langle x_1, \ldots, x_n \rangle \in W \}\,.
$$

\subsection{Multi-Tape Finite Automata}
A two-way finite-state automaton with $n \geq 1$ tapes scans $n$ read-only input tapes, each with an independent head.
At every step, the transition function determines the possible next states and head movements, based on the current state and the symbols currently under each head.
Two special symbols $\lmk, \rmk$ respectively mark the left and right ends of each input tape; $\Sigmaext$ denotes the extended alphabet $\Sigma\cup \{\lmk, \rmk\}$.


\begin{definition} \label{def:NDA}
A \emph{two-way nondeterministic finite-state automaton with $n$ tapes} is a tuple $\langle \Sigma, Q, \delta, q_0, F \rangle$, where:
\begin{itemize}
\item $\Sigma$ is the input alphabet, such that $\lmk, \rmk \not\in \Sigma$;
\item $Q$ is the finite set of states;
\item $\delta: Q \times \Sigmaext^n \to \powerset{Q \times \{-1, 0, 1\}^n}$ is the transition function that maps current state and input to a set of next states and head movement directions, with the restriction that the head does not move past the end markers;
\item $q_0 \in Q$ is the initial state;
\item $F \subseteq Q$ is the set of accepting states.
\end{itemize}
\end{definition}

The semantics of two-way nondeterministic finite-state automata relies on the notion of \emph{configuration}.
Given a two-way nondeterministic finite-state automaton $A$ as in Definition~\ref{def:NDA}, a configuration of $A$ is a $(2n+1)$-tuple 
$$
\langle x_1, \ldots, x_n, q, i_1, \ldots, i_n \rangle \ \in\ 
(\lmk \Sigma^* \rmk)^n \times Q \times \naturals^n\,,
$$ 
where $q \in Q$ is the current state, and, for $1 \leq k \leq n$, $x_k$ is the content of the $k$-th tape and $0 \leq i_k \leq |x_k| + 1$ is the position of the $k$-th head; when $i_k = 0$ (resp., $i_k = |x_k|+1$) the head is on the left marker $\lmk$ (resp., right marker $\rmk$).

The transition relation $\vdash$ between configurations is defined as:
$$\langle x_1, \ldots, x_n, q, i_1, \ldots, i_n \rangle \ \vdash\ \langle x_1, \ldots, x_n, q', i_1', \ldots, i_n' \rangle$$ 
if and only if, for each $1 \leq k \leq n$, $\sigma_{i_k,k}$ is the symbol at position $i_k$ in input word $x_k$, $\delta(q,\sigma_{i_1,1}, \ldots, \sigma_{i_n,n})$ includes a tuple $\langle q', d_1, \ldots, d_n \rangle$, and $i_k' = i_k + d_k$ for each $1 \leq k \leq n$.

A \emph{run} $\rho$ of $A$ on input $x = \langle x_1,\ldots,x_n \rangle$ is a sequence of configurations $\chi_0\,\ldots\,\chi_m$ such that $\chi_0 = \langle x, q_0, 0^n \rangle$ and, for all $1 \leq k \leq m$, $\chi_{k-1} \vdash \chi_k$.
A run $\rho$ of $A$ on input $x$ is \emph{accepting} if $\chi_m = \langle x, q, i_1, \ldots, i_n\rangle$ for some $q \in F$ and the $i_k$-th character of the $k$-th tape is $\rmk$ (that is, every head has reached the end of its tape).
Correspondingly, $A$ accepts an input word $x$ if there is an accepting run $\rho$ of $A$ on $x$.
The \emph{language} accepted (or recognized) by $A$ is the set of $n$-words
$$\lang{A} = \left\{\,x \in (\Sigma^*)^n \mid A \text{ accepts } x \,\right\}\,.$$

\begin{definition} \label{def:n-tape-automaton}
An $n$-tape automaton $A$ is:
\begin{itemize}
\item \emph{deterministic} if $|\delta(q, \sigma_1, \ldots, \sigma_n)| \leq 1$ for any $q, \sigma_1, \ldots, \sigma_n$;
\item \emph{$s$-synchronized} for $s \geq 0$ if every run of $A$, accepting or not, is such that any two heads that are not on the right-end marker $\rmk$ are no more than $s$ positions apart (as measured from the left-end marker $\lmk$);
\item \emph{synchronized} if it is $s$-synchronized for \emph{some} $s \in \naturals$;
\item \emph{asynchronous} if it is \emph{not} synchronized;
\item \emph{synchronous} if it is $0$-synchronized;
\item \emph{$r$-reversal bounded} for $r \geq 0$ if every run of $A$, accepting or not, is such that any head never inverts its direction of motion more than $r$ times;
\item \emph{reversal-bounded} if it is $r$-reversal bounded for \emph{some} $r \in \naturals$;
\item \emph{one-way} if it is $0$-reversal bounded, that is $\delta$ never moves any head left;
\item \emph{with rewind} if it only makes a special type of reversals where, after all heads reach the right-end marker $\rmk$, they are simultenously rewinded to the left-end marker $\lmk$, where the computation continues;
\item \emph{$r$-rewind bounded} for $r \geq 0$ if it is $2r$-reversal bounded with rewind;
\item \emph{rewind-bounded} if it is $r$-rewind bounded for \emph{some} $r \in \naturals$.
\end{itemize}

Correspondingly, $\nfa{n,s,r}$ denotes the class of $s$-synchronized $r$-reversal \linebreak bounded $n$-tape two-way nondeterministic finite-state automata, $\dfa{n,s,r}$ the corresponding deterministic class, and when $s = \infty$ (resp.\ $r = \infty$) the class includes asynchronous (resp.\ two-way reversal-unbounded) automata as well.
Finally, for $\ift{r} \in \naturals \cup \{\infty\}$, $\nfa{n,s,\rb{\ift{r}}}$ and $\dfa{n,s,\rb{\ift{r}}}$ denote the classes of nondeterministic and deterministic $s$-synchronized $\ift{r}$-rewind bounded (or unbounded if $\ift{r} = \infty$) $n$-tape two-way finite-state automata.

With a little abuse of notation, the notation for automata classes also denotes to the corresponding class of accepted languages; for example, $\nfal{3,\infty,4}$ is also the class of languages recognized by nondeterministic 3-tape machines with at most 4 reversals and with no synchronization restrictions.
\end{definition}

\subsection{Weak Inclusions}
A number of weak inclusions are a direct consequence of the definitions.
In the following, let $n \geq 1$ be a number of tapes, $s,r \in \naturals$ be nonnegative integers, and $\ift{s},\ift{r} \in \naturals \cup \{\infty\}$ be nonnegative integers or infinity.
Determinism is a syntactic restriction, thus:
\begin{equation}
\dfal{n,\ift{s},\ift{r}} \subseteq \nfal{n,\ift{s},\ift{r}}\,;
\end{equation}
and similarly, bounding the number of reversals reduces generality:
\begin{gather}
\dfal{n,\ift{s},r} \subseteq \dfal{n,\ift{s},r+1} \subseteq \dfal{n,\ift{s},\infty}\,, \\
\nfal{n,\ift{s},r} \subseteq \nfal{n,\ift{s},r+1} \subseteq \nfal{n,\ift{s},\infty}\,;
\end{gather}
and so does limiting the degree of synchrony:
\begin{gather}
\dfal{n,s,\ift{r}} \subseteq \dfal{n,s+1,\ift{r}} \subseteq \dfal{n,\infty,\ift{r}}\,, \\
\nfal{n,s,\ift{r}} \subseteq \nfal{n,s+1,\ift{r}} \subseteq \nfal{n,\infty,\ift{r}}\,.
\end{gather}
Rewinds are a restricted form of reversals, hence:
\begin{gather}
\dfal{n,\ift{s},\rb{\ift{r}}} \subseteq \dfal{n,\ift{s},2\ift{r}}\,, \\
\nfal{n,\ift{s},\rb{\ift{r}}} \subseteq \nfal{n,\ift{s},2\ift{r}}\,.
\end{gather}

\subsection{Decision Problems}
\label{sec:decision-problems}

Given a class $\mathcal{C}$ of languages, and $n$-tape automata $A,B$, consider the following decision problems:
\begin{description}
\item[Emptiness:] does $\lang{A} = \emptyset$?
\item[Universality:] does $\lang{A} = (\Sigma^*)^n$?
\item[Finiteness:] is $\lang{A}$ finite?
\item[Disjointness:] does $\lang{A} \cap \lang{B} = \emptyset$?
\item[Inclusion:] does $\lang{A} \subseteq \lang{B}$?
\item[Equivalence:] does $\lang{A} = \lang{B}$?
\item[Class membership:] does $\lang{A} \in \mathcal{C}$?
\end{description}

\subsection{Multi-Tape Automata and the Rational Languages}
\label{sec:rational-terminology}

The languages accepted by variants of multi-tape automata (Definition~\ref{def:n-tape-automaton}) have also been studied by algebraic means: see for example \cite[Chap.~3]{Berstel79-book} and \cite[Chap.~4]{Saka09-book}.
It is convenient to be aware of the terminology used in those works:
\begin{itemize}
\item The class \nfa{n, \infty, 0} of languages accepted by nondeterministic one-way automata without synchronization requirements is called the \emph{rational languages}.
\item For $s$ a nonnegative integer, the class \nfa{n, s, 0} of languages accepted by nondeterministic one-way $s$-synchronized automata is called the rational languages \emph{with delay (or lag) $s$}.
\item The class \nfa{n, 0, 0} of languages accepted by nondeterministic one-way synchronous automata is called the \emph{automatic languages}.
\end{itemize}

\section{Single-Tape Automata} \label{sec:single-tape-automata}
This section recalls the well-known properties of single-tape automata
(that is, with $n=1$ tapes), which define the class of \emph{regular}
languages~\cite{RabinScott59,HMU06,Kozen-automata}.  Clearly, the
notion of synchronization is immaterial with a single tape:
\begin{equation}
\dfa{1,0,0} = \nfa{1,0,0} = \dfa{1,\ift{s},\ift{r}} = \nfa{1,\ift{s}',\ift{r}'}\,.
\end{equation}
Also, languages defined by one-tape automata are closed under complement, intersection, union, concatenation, Kleene closure, projection, generalization, and reversal.
The emptiness, universality, finiteness, disjointness, inclusion, equivalence, and class membership problems are all decidable for regular languages.

\section{Synchronized Multi-Tape Automata} \label{sec:synchr-multi-tape}
This section studies several properties of synchronized multi-tape automata $\nfa{n,s,\ift{r}}$ and $\dfa{n,s,\ift{r}}$, where $s$ denotes a generic value in $\naturals$, and $\ift{r}$ a generic value in $\naturals \cup \{\infty\}$.
As we will see, synchronization essentially makes multi-tape automata defining regular languages over convolutions, hence synchronized multi-tape automata define a very robust class of languages.

\subsection{Expressiveness}
This section shows that: 
\begin{equation}
  \dfa{n,0,0} = \nfa{n,0,0} = \dfa{n,s,\ift{r}} = \nfa{n,s,\ift{r}}\,.
\label{eq:synchronized}
\end{equation}
Since rewinds are a special case of reversals, it also follows that:
\begin{equation}
  \dfa{n,0,0} = \nfa{n,0,0} = \dfa{n,s,\rb{\ift{r}}} = \nfa{n,s,\rb{\ift{r}}}\,.
\label{eq:synchronized-rewind}
\end{equation}

\subsubsection{Increasing Synchronization}
Every synchronized $n$-tape automaton can be transformed to an equivalent synchronous $n$-tape automaton:
\begin{equation}
\nfa{n,s,\ift{r}} = \nfa{n,0,\infty}\,.
\label{eq:to-0-sync}
\end{equation}
The construction to translate an $A \in \nfa{n,s,\ift{r}}$ into an $A' \in \nfa{n,0,\ift{r}}$ generalizes the one in \cite{citeulike:10289197} (\cite{FrougnyS93} proved the same result by algebraic means) to two-way nondeterministic automata; to our knowledge, it is original to the present survey.

It works as follows: additional states in the synchronous automaton $A'$ keep track of the ``neighbour'' $s$ symbols on each tape scanned during a synchronous reading of the tapes.

Precisely, the states $Q'$ of $A'$ are of the form $[q, b_1, \ldots, b_n]$, where each $b_k$ is a word of at most $s+1$ characters over $\Sigmaext \cup \{\mk\}$, with exactly one occurrence of the special symbol $\mk$.
For any such words $b$, $\mkat{b}$ denotes the character immediately to the right of $\mk$, or $\epsilon$ if $\mk$ is the right-most character in $b$.

Then, a configuration of $A'$ where the symbol currently under the $k$-th head is $\sigma_k$ and the current state is $[q,b_1, \ldots, b_n]$, corresponds to a configuration of $A$ where the symbol currently under the $k$-th head is $\mkat{b_k \cdot \sigma_k}$---that is the one immediately to the right of $\mk$ in $b_k \cdot \sigma_k$---and the current state is $q$.

The initial state $q_0'$ of $A'$ is $[q_0, \mk^n]$, where all heads are on the left marker.
During any computation over a given input, the (common) position of the heads in $A'$ corresponds to the position of the right-most head in $A$; more precisely, the states of $A'$ plus the input describe a $n \times (k+1)$ tape portion:
\begin{equation*}
\begin{array}{llllll}
\alpha_1 & x_1 & \mk & y_1 & \beta_1 & \sigma_1' \\
\vdots & \vdots &\vdots & \vdots & \vdots & \vdots \\
\alpha_k & x_k & \mk & y_k & \beta_k & \sigma_k' \\
\vdots & \vdots &\vdots & \vdots & \vdots & \vdots \\
\alpha_n & x_n & \mk & y_n & \beta_n & \sigma_n' \,,
\end{array}
\end{equation*}
where $\alpha_k \in \Sigmaext^*$ is the left-most stored portion of the $k$-th tape, $x_k \in \{\epsilon\} \cup \Sigmaext$ is the character immediately to the left of the marker, $y_k \in \{\epsilon\} \cup \Sigmaext$ is the character immediately to the right of the marker, $\beta_k \in \Sigmaext^*$ is the right-most stored portion of the $k$-th tape, and $\sigma_k' \in \Sigmaext$ is the charater currently under the $k$-th head.
The hypothesis that $A'$ is $s$-bounded entails that $2 \leq |\alpha_k \, x_k \, \mk \, y_k \, \beta_k \, \sigma_k'| \leq s+2$.
Finally, when a head is on the right marker $\rmk$, we call it ``inactive'', and we do not update the corresponding portion of the state.
Write $\inact$ to denote the set of inactive heads in the current state (that is, the set of heads $h$ such that $a_h \cdot \sigma_h \in \Sigma^* \rmk$ and $a_hx_h \neq \epsilon$), and $\act = \{1, \ldots, n\} \setminus \inact$ to denote the other active heads.

Let us now see how to construct the transitions of $A'$. For each of $A$'s transitions 
$$\langle q', d_1, \ldots, d_n\rangle \in \delta(q, \sigma_1, \ldots, \sigma_n)\,,$$
consider all states of $A'$ of the form $[q, a_1, \ldots, a_n]$.
$A'$ should make a transition from such current state according to the characters currently marked by $\mk$, corresponding to the $\sigma_i$, and then adjust the state to reflect the head movements in $A$.
If all \emph{active} heads move left in $A$ (that is, $\max \{d_j \mid j \in \act\} = -1$) and there exist an active tape $\overline{k}$ whose $s$-size buffer encoded in the state has the marker $\mk$ in the leftmost position ($\alpha_{\overline{k}} = x_{\overline{k}} = \epsilon$), then it is not sufficient to rearrange the information in $[q, a_1, \ldots, a_n]$ to determine the next state $A'$ moves to, because the new character the head moves is to $s+1$ characters away from the current position of the head in $A'$.
Hence, in these situations, $A'$ enters a sequence of \emph{lookup} states $\ell_1^-, \ldots, \ell_{s+1}^-$ while moving all its active heads left, until it reads the $(s+1)$-th character to the left, so that it can update the state accordingly.
Precisely, the following transitions in $A'$ define a lookup, for any characters $\mu_1, \ldots, \mu_n$ read while moving heads.
\begin{itemize}
\item Enter the lookup state $\ell_1^-$ and start moving left:
$$
\langle [\ell_1^-,a_1, \ldots, a_n], \lambda_1^{-1/0}, \ldots, \lambda_n^{-1/0} \rangle \in \delta'([q, a_1, \ldots, a_n], \mu_1, \ldots, \mu_n)\,,
$$
where 
$$
\lambda_j^{a/b} \ = \ 
\begin{cases}
  a  &  j \in \act\,, \\
  b   &  \text{otherwise}\,,
\end{cases}
$$
hence inactive heads are not moved.
\item Traverse all lookup states, for $1 \leq h \leq s$, while moving left:
$$
\langle [\ell_{h+1}^-,a_1, \ldots, a_n], \lambda_1^{-1/0}, \ldots, \lambda_n^{-1/0} \rangle \in \delta'([\ell_h^-, a_1, \ldots, a_n], \mu_1, \ldots, \mu_n)
$$
\item Include the current characters $\mu_1, \ldots, \mu_n$ in the state and start moving right on the ``duplicate'' lookup state $\ell_s^+$:
$$
\langle [\ell_{s}^+, \nu_1, \ldots, \nu_n], \lambda_1^{1/0}, \ldots, \lambda_n^{1/0} \rangle \in \delta'([\ell_{s+1}^-, a_1, \ldots, a_n], \mu_1, \ldots, \mu_n)\,,
$$
where $\nu_k$ is $\lambda_k^{\mu_k/\epsilon} \cdot a_k$, with the last character removed if $|\lambda_k^{\mu_k/\epsilon} \cdot a_k| > s+1$.
\item Move right, for $s \geq h > 1$, until you reach $\ell_1^+$:
$$
\langle [\ell_{h-1}^+,\nu_1, \ldots, \nu_n], \lambda_1^{1/0}, \ldots, \lambda_n^{1/0} \rangle \in \delta'([\ell_h^+, \nu_1, \ldots, \nu_n], \mu_1, \ldots, \mu_n)
$$
\item Move to the original position on $q$:
$$
\langle [q,\nu_1, \ldots, \nu_n], \lambda_1^{1/0}, \ldots, \lambda_n^{1/0} \rangle \in \delta'([\ell_1^+, \nu_1, \ldots, \nu_n], \mu_1, \ldots, \mu_n)\,.
$$
Notice that, after the lookup, the tuple $\nu_1,\ldots, \nu_n$ is certainly different than the starting tuple $a_1, \ldots, a_n$, because we have added a left-most character to the left of the marker $\mk$ in at least one of $a_1, \ldots, a_n$.
Hence, the lookup does not introduce nondeterminism.
\end{itemize}

After setting up the lookup when necessary, let us now describe normal transitions of $A'$.
Let $[q, a_1, \ldots, a_n]$ be the current state; add to $A'$ all transitions of the form 
$$\langle [q', a_1', \ldots, a_n'], \lambda_1^{d'/0}, \ldots, \lambda_n^{d'/0} \rangle \in \delta'([q, a_1, \ldots, a_n], \sigma_1', \ldots, \sigma_n')$$ 
such that, for $1 \leq k \leq n$:
\begin{itemize}
\item $d' = \max \{d_j \mid j \in \act \}$;
\item $\sigma_k = \mkat{a_k \cdot \sigma_k'}$ or, if $\mk$ does not appear in $a_k$ (this may only happen after a lookup when the $\mk$ occurred at the last position in $a_k$), $\sigma_k = \sigma_k'$;
\item Let $\gamma_k$ be $a_k \cdot \sigma_k'$ with $\mk$ moved right, left, or not moved, according to whether $d_k$ is $1$, $-1$, or $0$; formally:
\begin{equation*}
\gamma_k \ = \ 
\begin{cases}
\alpha_k \, x_k\, y_k \mk\beta_k\, \sigma_k'  &  d_k = 1 \text{ and } y_k \neq \epsilon\,, \\
\alpha_k \, x_k \, \sigma_k' \mk  &  d_k = 1 \text{ and } y_k = \epsilon= \beta_k\,, \\
\alpha_k \, x_k \mk y_k \,\beta_k\, \sigma_k'  &  d_k = 0\,,  \\
\alpha_k \mk x_k\, y_k \,\beta_k\, \sigma_k'  &  d_k = -1 \text{ and } \mk \text{ appears in } a_k\cdot\sigma_k' \,, \\
a_k \mk \sigma_k'  &  d_k = -1 \text{ and } \mk \text{ doesn't appear in } a_k\cdot\sigma_k' \,.
\end{cases}
\end{equation*}
Then, $a_k'$ is:
\begin{equation*}
a_k' \ = \ 
\begin{cases}
\gamma_k  &  |\gamma_k| \leq s+1 \text{ and } d' = 1\,, \\
\gamma_k \text{ without the first character } &  |\gamma_k| > s+1 \text{ and } d' = 1 \,, \\
\gamma_k \text{ without the last character } \sigma_k' &  d' \leq 0 \,. \\
\end{cases}
\end{equation*}
\end{itemize}

\paragraph{Remarks.}
\begin{itemize}
\item With this construction the number $|Q'|$ of states of $A'$ is $\mathrm{O}(|Q|\cdot|\Sigma|^{n\cdot s} \cdot s |\Sigma|)$, because there are at most $|\Sigma|^{n\cdot s}$ $n$-tuples of words of length at most $s$ for each state in $Q$, plus a multiplicative factor $s\cdot |\Sigma|$ to account for the lookup states ($2s + 1$, for any possible left-most characters $\mu_k \in \Sigmaext$).

\item The lookups may introduce additional reversals, therefore the construction does not, in general, preserve the original value $\ift{r}$ of reversals.

\item Automaton $A'$ is deterministic if and only if $A$ is. Therefore, the same construction proves:
\begin{equation}
\dfa{n,s,\ift{r}} = \dfa{n,0,\infty}\,.
\label{eq:to-0-sync-det}
\end{equation}
\end{itemize}

\subsubsection{Removing Reversals}
The classic constructions to turn two-way automata into equivalent one-way automata (such as Shepherdson's~\cite{Shep-2wayautomata} or Vardi's~\cite{Vardi-2wayautomata}) are applicable to multi-tape synchronous automata as well.
The only minimal difference is that a multi-tape one-way automaton is not forced to always move its heads right, because the heads reaching the right-hand marker $\rmk$ cannot move right at all. 
In all, the following result holds:
\begin{equation}
\nfa{n,0,\ift{r}} = \nfa{n,0,0}\,.
\label{eq:to-one-way}
\end{equation}

\subsubsection{Determinization}
The usual construction to determinize nondeterministic finite-state automata works for multi-tape synchronous automata too:
\begin{equation}
\dfa{n,0,0} = \nfa{n,0,0}\,.
\label{eq:to-deterministic}
\end{equation}

\subsubsection{Convolution}\label{sec:sync-convolution}
The computations of synchronous $n$-tape one-way automata can be regarded as computations of single-tape automata over $n$-track alphabets.
More precisely, let $\lang{A}$ be the language over alphabet $\Sigma$ accepted by some synchronous $n$-tape one-way automaton $A \in \nfa{n,0,0}$.
Then, there exists a single-tape one-way automaton $B$ that accepts the language $\convof{\lang{A}}$.

\subsection{Closure Properties}
Since synchronized automata essentially define regular languages via convolution, they enjoy the same closure properties.
Thus, the usual constructions~\cite{HMU06} show that \linebreak $\nfa{n,0,0}$ is closed under under complement, intersection, union, concatenation, Kleene closure, projection, generalization, and reversal.

\subsection{Decidability}
Decidability of most decision problems considered in Section~\ref{sec:decision-problems} is also a straightforward consequence of the regular nature of synchronized automata: for example, \emph{emptiness} is decidable: construct the automaton recognizing $\convof{\lang{A}}$ and determine if it accepts some word; alternatively, directly test reachability of an accepting state.
Similarly, it is easy to see that universality, finiteness, disjointness, inclusion, and equivalence are decidable for synchronized automata.

Next, we consider some class membership problems.
Ibarra and Tran~\cite[Th.~15]{citeulike:10289197} show that, for generic two-way automaton $A \in \nfa{n,\infty, \infty}$ and $s \geq 0$, it is decidable to determine whether $A \in \nfa{n,s,\infty}$, that is whether $A$ is $s$-synchronized.
The idea is to build an automaton $M$ that simulates the computations of $A$ on a single tape that stores the convolution of the input; whenever $M$ detects a computation where a pair of heads would ``separate'' more than $s$ cells apart, it accepts, and otherwise it rejects.
Since $M$ accepts the empty language if and only if $A$ is not $s$-synchronized, and emptiness for $M$ is decidable (as $M$ is a single-tape automaton), it follows that we have an effective procedure to determine whether $A \in \nfa{n,s,\infty}$.
The same procedure is applicable to automata with bounded reversals, hence it is also decidable whether $A \in \nfa{n,s,\ift{r}}$ for given $s \in \naturals$ and $\ift{r} \in \naturals \cup \{\infty\}$.

Ibarra and Tran~\cite[Th.~14]{citeulike:10289197} also show---via reduction from the halting problem of 2-counter machines---that, for a generic two-way automaton $A \in \nfa{n,\infty, \infty}$, it is undecidable to determine whether $A$ is synchronized (for some $s \in \naturals$).
However, if $A \in \dfa{n,\infty, 0}$ is one-way deterministic, then the same problem is decidable, because $A$ is asynchronous unless it is $(q-1)$-synchronized, where $q$ is the number of $A$'s states.
Since we notice that the determinization construction of single-tape automata is applicable to deterministic one-way multi-tape automata as well, we have that the problem of deciding whether a generic $A \in \nfa{n,\infty, 0}$ is synchronized is also decidable.

On the other hand, the following problem is undecidable (for $n > 1$)~\cite[Th.~11]{citeulike:10289197}: given $A \in \nfa{n,\infty,0}$, determine whether $\lang{A} \in \nfa{n,s,0}$ with $s$ given or not (the two problems are both undecidable because of the result \eqref{eq:synchronized}: every $s$-synchronous language is also $0$-synchronous), that is whether $A$ recognizes a synchronous language.
Undecidability immediately extends to the two-way case (also with bounded reversals).
The decidability of the corresponding problems for deterministic one-way automata is currently open.

\section{Asynchronous Multi-Tape Automata} \label{sec:asynchr-multi-tape}
This section studies several properties of asynchronous multi-tape automata $\nfa{n,\infty,\ift{r}}$ and $\dfa{n,\infty,\ift{r}}$, where $\ift{r}$ denotes a generic value in $\naturals \cup \{\infty\}$, and $n \geq 2$.
Asynchrony significantly increases the expressiveness of multi-tape automata, but it also significantly restricts the decidability and closure properties.

\subsection{Expressiveness}
This section shows that: 
\begin{gather}
  \nfa{n,s,\ift{r}} \subset \dfa{n,\infty,0} \subseteq \dfa{n,\infty,\rb{r}} \subset \dfa{n,\infty,\rb{r+1}}
\,,
\label{eq:aynchronous-rewind}
\\
\dfa{n,\infty,\rb{r}} \subset \nfa{n,\infty,\rb{r}} \subset \nfa{n,\infty,\rb{r+1}}\,,
\label{eq:aynchronous-rewind-nondet}
\\
\dfa{n,\infty,\rb{0}} = \dfa{n,\infty,0} \subset \nfa{n,\infty,0} = \nfa{n,\infty,\rb{0}} \,,
\label{eq:aynchronous-reversal}
\end{gather}
and, for $r \geq 1$,
\begin{gather}
\nfa{n,\infty,\rb{r}} \subset \dfa{n,\infty,2r} \subseteq \nfa{n,\infty,2r}\,,
\label{eq:aynchronous-reversal}
\\
\nfa{n,\infty,0} \text{ and } \dfa{n,\infty,\rb{r}}  \text{ are incomparable.}
\label{eq:nondet-rewind-incomparable}
\end{gather}

\subsubsection{Asynchronous More Expressive Than Synchronous}
There exist asynchronous languages that no synchronous automaton recognizes, thus:
\begin{equation}
\nfa{n,s,\ift{r}} = \dfa{n,s,\ift{r}} \subset \dfa{n,\infty,0}\,.
\end{equation}
In fact, consider the language over pairs of strings
$$
L_{n,2n} \ =\  \{ \langle a^n, a^{2n} \rangle \mid n \in \naturals \}\,.
$$
It is clear that $L_{n,2n} \in \dfa{n,\infty,0}$, because the automaton in Figure~\ref{fig:Ln2n} recognizes it; notice that the automaton is indeed asynchronous, as the second head makes twice as many moves as the first one, and both scan the entire tape upon acceptance.
On the contrary, 
$$
L_{n,2n}^{\otimes} \ =\ 
\left\{
\left[
\begin{array}{c}
a \\ a
\end{array}
\right]^n
\left[
\begin{array}{c}
\pad \\ a
\end{array}
\right]^n
\mid
n \in \naturals
\right\}
$$
is not regular, hence $L_{n,2n}$ is not accepted by any synchronous $2$-tape automaton because of the results of Section~\ref{sec:sync-convolution}.
Since we can always extend $L_{n,2n}$ to an $n$-word language with all components empty but the first two, the separation of synchronized and asynchronous holds for every $n \geq 2$.

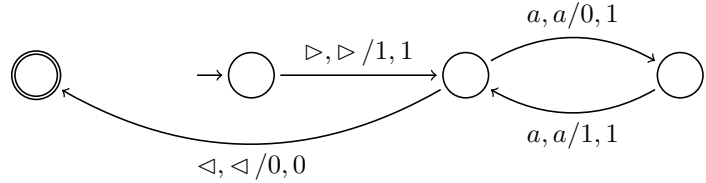
\begin{figure}[!hb]
\centering
\begin{tikzpicture}[->,semithick, accepting/.style=accepting by double,initial text=,
  every state/.style={draw,minimum size=6mm,inner sep=1pt},
  shorten >=2pt, shorten <=2pt, node distance=22mm]
\node [state,initial] (init) {};
\node [state] (read2n) [right=of init] {};
\node [state] (readn) [right=of read2n] {};
\node [state,accepting] (done) [left=of init] {};

\path (init) edge node [above] {$\lmk,\lmk/1,1$} (read2n);
\path (read2n) edge [bend left] node [above] {$a,a/0,1$} (readn);
\path (readn) edge [bend left] node [below] {$a,a/1,1$} (read2n);

\path (read2n) edge [bend left] node [below] {$\rmk,\rmk/0,0$} (done);
\end{tikzpicture}
\caption{A 2-tape asynchronous automaton recognizing the language $L_{n,2n}$.}
\label{fig:Ln2n}
\end{figure}

\subsubsection{Expressiveness Increases With Rewinds}
\label{sec:expr-incr-with-rewinds}
The power of asynchronous multi-tape automata increases with the number of rewinds:
\begin{gather}
\dfa{n,\infty,\rb{r}} \subset \dfa{n,\infty,\rb{r+1}}\,, \\
\nfa{n,\infty,\rb{r}} \subset \nfa{n,\infty,\rb{r+1}}\,.
\end{gather}
Collectively, rewind-bounded automata are instead as expressive as rewind-unbounded ones:
\begin{gather}
\bigcup_{r \in \naturals} \dfa{n,\infty,\rb{r}} = \dfa{n,\infty,\rb{\infty}}\,, 
\label{eq:rewind-bounded-det}
\\
\bigcup_{r \in \naturals} \nfa{n,\infty,\rb{r}} = \nfa{n,\infty,\rb{\infty}}\,.
\label{eq:rewind-bounded-nondet}
\end{gather}

These results have been proved by Chan~\cite[Ch.~4]{Chan-PhD80}.
Consider first deterministic rewind-bounded automata.
Given any $2^r$ distinct positive integers $$0 < k_1 < k_2 < \cdots < k_{2^r}\,,$$ consider the language
$$
L_{2^r} \ =\ \{ \langle a^m,a^{k_i\cdot m} \rangle \mid m \in \naturals, 1 \leq i \leq 2^r \} \,.
$$
It is not difficult to show that $L_{2^r} \in \dfa{2,\infty,\rb{r}}$: each language $\{\langle a^m,a^{k \cdot m} \rangle \mid \linebreak m \in \naturals\}$ is recognizable without rewinds, and the $r$ rewinds can implement a binary search among all $2^r$ different values of $k_1, k_2, \ldots, k_{2^r}$.
The standard argument that binary search of $2^r$ elements requires, in the worst case, $r$ comparisons, carries over to this setting of $r$-rewind automata, hence allowing us to prove that $r$ rewinds are necessary to recognize $L_{2^r}$.
The argument clearly extends to an arbitrary number of tapes, thus separating $\dfa{n,\infty,\rb{r}}$ from $\dfa{n,\infty,\rb{r+1}}$ for all finite $r$'s.

Let us now consider nondeterministic rewind-bounded automata.
The argument for deterministic ones breaks down, as we can use nondeterministic parallelism to try out all $2^r$ values of $k$.
More generally, every language over \emph{unary} alphabet that is accepted by a reversal-bounded $n$-tape automaton is also accepted by some one-way $n$-tape nondeterministic automaton (that can be constructed effectively, see Chan~\cite[Th.~4.1]{Chan-PhD80}).
However, inclusion is still strict for larger alphabets.
For $m \geq 2$, consider the $2$-word language
$$
L_{0/1,m} \ =\ 
\left\{
\langle
x_1 \# x_2 \# \cdots \# x_m ,
x_m \# x_{m-1} \# \cdots \# x_1
\rangle
\mid
x_1, \ldots, x_m \in \{0,1\}^*
\right\}\,.
$$
It is clear that $L_{0/1,m}$ is accepted by a deterministic $(m-1)$-rewind bounded automaton.
Chan proves, using Yao and Rivest's technique~\cite{YaoR78}, that no nondeterministic automaton with only $m-2$ rewinds can accept $L_{0/1,m}$.
Since a deterministic automaton is a special case of a nondeterministic one, we have a separation of $\nfa{2,\infty, m-2}$ and $\nfa{2,\infty, m-1}$ (which, as usual, extends to a generic number of tapes).

Finally, Rosenberg first observed~\cite{Rosenberg67} that bounding the number of rewinds is without loss of generality for deterministic automata: if, in some computation, an automaton re-enters the same state after two rewinds, it has entered an infinite loop (and hence will not accept the input).
The pigeonhole principle implies that an automaton must re-enter the same state after a number of rewinds equal to the number of its internal states.
Thus, a finite bound on the number of rewinds (equal to the number of internal states) does not limit the expressiveness of deterministic automata.
A very similar argument works for nondeterministic automata too (as observed by Chan): an input can be accepted only if there exists some computation where a different state is entered after each rewind; therefore, every accepting computation has bounded rewinds.
These arguments prove \eqref{eq:rewind-bounded-det} and \eqref{eq:rewind-bounded-nondet}.

\subsubsection{Expressiveness and Reversals}
It is currently unknown whether a hierarchy on to the number of reversals exists, similar to the hierarchy on the number of rewinds discussed in the previous section.
In particular, it is unknown whether the inclusions $\dfa{n,\infty,r} \subseteq \dfa{n,\infty,r+1}$ and $\nfa{n,\infty,r} \subseteq \nfa{n,\infty,r+1}$ are strict for every $r$.
The standard conjecture is that they are, but the capabilities of reversals are quite difficult to capture exactly for asynchronous automata.
In particular, Chan~\cite[Sec.~4.3]{Chan-PhD80} showed that the hierarchy collapses for nondeterministic automata over unary alphabet, since these automata precisely define the set of unary encodings of Presburger relations, regardless of the number of reversals; thus, a separation in the general case requires languages over binary alphabets.

\subsubsection{Nondeterminism Not Replaceable by Rewinds} \label{sec:nond-not-repl-rewinds}
Nondeterminism is not replaceable by rewinds in general.
Consider the language $L_{2^r}$ introduced in Section~\ref{sec:expr-incr-with-rewinds}; a deterministic automaton needs $r$ rewinds to recognize it, but it is clear that a nondeterministic automaton needs no rewinds, as it can just guess the right value of $k$ (if it exists).
This shows that $\nfa{n, \infty, 0} \setminus \dfa{n, \infty, \rb{r}}$ is not empty, for each $r \geq 0$ and $n \geq 2$.

Chan~\cite[Th.~4.6]{Chan-PhD80} proves a stronger result: there exist languages accepted by one-way nondeterministic automata that no deterministic automaton with (arbitrarily many) rewinds accepts.
The language
$$
L_{\star,m} \ =\ 
\left\{
\left\langle\begin{array}{l}
u_1\star v_1 \# \cdots \# u_m\star v_m ,\\
x_1\star y_1 \# \cdots \# x_n\star y_n
\end{array}\right\rangle
\left|
\begin{array}{c}
\left(
\begin{array}{l}
u_1, \ldots, u_m, \\ 
v_1, \ldots, v_m, \\
x_1, \ldots, x_n, \\
y_1, \ldots, y_n 
\end{array}
\right)
\in \{0,1\}^+\,, \\
\exists i,j: u_i = x_j \;\wedge\; v_i \neq y_j
\end{array}
\right.
\right\}
$$
is an example that belongs to the non-empty difference
\begin{equation}
\nfa{n,\infty,0} \setminus \bigcup_{r \in \naturals} \dfa{n,\infty,\rb{r}}\,.
\label{eq:rewind-bounded-detminusnondet}
\end{equation}

Rosenberg~\cite[Th.~9]{Rosenberg67} shows a simpler language that belongs to the set difference \eqref{eq:rewind-bounded-detminusnondet}.
Consider the two languages 
\begin{align*}
B \quad & =\quad \{\langle xa,\epsilon \rangle \mid x \in \{a,b\}^*\}\,, \\
C \quad & =\quad \{\langle b^m,b^m \rangle \mid m \in \naturals \}\,,
\end{align*}
and their concatenation 
\begin{align*}
A = B \cat C & = \quad 
\{\langle xab^m,b^m \rangle \mid m \in \naturals, x \in \{a,b\}^*\}
\,.
\end{align*}
It is clear that both $B$ and $C$ are in $\dfa{2,\infty,0}$.
On the contrary, $A$ is not in $\dfa{2,\infty,\rb{r}}$ for any $r \geq 0$ because a deterministic automaton cannot ``guess'' where the tail of $b$'s starts on the first tape; rewinds do not help, because the tail can be arbitrarily long, while the number of rewinds is bounded by the finite number of states of the automaton (as we observed in Section~\ref{sec:expr-incr-with-rewinds}).
For the same reason, $A$ is in $\nfa{2,\infty,0}$, where a nondeterministic choice guesses where the tail starts.

\subsubsection{Rewinds Not Replaceable by Nondeterminism}
Rewinds are not replaceable by nondeterminism in general, as first proved by Rosenberg~\cite[Th.~9]{Rosenberg67}.
We can easily show it using the language 
$$
L_{m,m} \{ \langle a^m b a^m, a^m b a^m \rangle \mid m \in \naturals\}\,,
$$
discussed later in Section~\ref{sec:det-one-way-automata-closures}.
$L_{m,m}$ is in $\dfa{2,\infty,\rb{1}}$: a deterministic automaton can make a pass to check that the two input strings are identical and in the form $a^*ba^*$; then, it rewinds and makes another pass to check that the sequence of $a$'s before the $b$ on the first tape has the same length as the sequence of $a$'s after the $b$ on the second tape.
However, $L_{m,m}$ is not in $\nfa{2,\infty,0}$, because nondeterminism does not help here (Section~\ref{sec:det-one-way-automata-closures} gives a more rigorous characterization of why this is the case).
Hence, the difference
\begin{equation}
\bigcup_{r \in \naturals} \dfa{n,\infty,\rb{r}} \setminus \nfa{n,\infty,0}
\label{eq:rewind-bounded-detvsnondet}
\end{equation}
is not empty.

\subsubsection{Reversals Not Replaceable by Rewinds}
Reversals are not replaceable by rewinds in general, not even when combined with nondeterminism.
The 2-word language
$$
L_{\rho} \ =\ \{ \langle x, \rev{x}\rangle \mid x \in \{0,1,\#\}^* \}
$$
is clearly recognizable with a 1-reversal bounded deterministic automaton.
No automaton with rewinds and nondeterminism, however, can accept $L_{\rho}$.
Otherwise, let $M$ be such an automaton, and let $\mu$ be the number of its rewinds; we could then use $M$ to recognize a language similar to $L_{0/1,m}$ for $m = \mu + 2$, but where the word on the second tape is the reversal of the one on the first; this is impossible with $\mu$ rewinds, as shown by same argument that showed that $L_{0/1,m}$ is not accepted with less than $m-1$ rewinds.
The contradiction shows that the difference $\dfa{n,\infty,1} \setminus \nfa{n,\infty,\rb{\ift{r}}}$ is not empty.

\subsubsection{Determinism vs.\ Nondeterminism}
Whether reversals can replace nondeterminism is an open problem in general; in particular, it is unknown whether the inclusion $\dfa{n,\infty,\ift{r}} \subseteq \nfa{n,\infty,\ift{r}}$ is strict or not.
Some related results for single-tape multi-head automata might suggest that it is.
For example, Bebj{\'a}k and {\v S}tef{\'a}nekov{\'a}~\cite{Bebjak88} show some functions $f(m)$ (where $m$ is the length of the input) such that $f(m)$-reversal bounded $n$-head nondeterministic automata are strictly more expressive than their deterministic counterparts.
The general question for $n$-head automata is likely hard to settle, as it corresponds~\cite{abs-0906-3051} to the open problem of whether deterministic logarithmic space (L) equals nondeterministic logarithmic space (NL).

\subsection{Closure Properties}
This section presents the closure properties of multi-tape finite automata.
We start with one-way automata, and then introduce rewinds and reversals.

\subsubsection{Deterministic One-Way Automata}
\label{sec:det-one-way-automata-closures}
Let us consider the closure properties of the class $\dt{n} = \dfa{n,\infty,0}$, for $n \geq 2$.

Rabin and Scott observed~\cite[Th.~17]{RabinScott59} that $\dt{2}$ is closed under complement, with the usual construction that complements the accepting states (with a transition function that is total).
The same construction carries over to show that $\dt{n}$ is closed under complement for any $n$.

$\dt{n}$ is not closed under projection, but we have the weaker property that if $L \in \dt{n}$ then $\proj{L}{k} \in \ndt{n-1}$.
Rabin and Scott proved this~\cite[Th.~16]{RabinScott59} for $n=2$, and other authors~\cite{ElgotMezei65,FischerR68,Monks10} generalized it.
The idea of the proof is the following: given an $n$-tape automaton $A$, build an $n-1$-tape automaton $B$ that replicates $A$'s behavior on all tapes except the $k$-th (which is projected out), where it behaves nondeterministically (that is, it performs any computation $A$ may perform on the $k$-th tape).

Since $\dt{n}$ is closed under complement but not under projection, it is also not closed under generalization: the projection $\proj{L}{k}$ of a language $L \in \dt{n}$ is equivalently expressed as $\comp{\fall{\comp{L}}{k}}$, hence if $\dt{n}$ were closed under generalization, it would also be closed under projection---a contradiction.

$\dt{n}$ is not closed under intersection.
For example, Rabin and Scott~\cite[Th.~17]{RabinScott59} suggest the $2$-word language:
\begin{equation*}
\begin{split}
L_{m,m} & =
\{ \langle a^m b a^m, a^m b a^m \rangle \mid m \in \naturals\} \\
&= L_{m} \cap L_{x,x} \ = \ 
\{\langle a^mba^h, a^kba^m \rangle \mid m,h,k \in \naturals \}
\cap
\{\langle x,x \rangle \mid x \in \{a,b\}^* \}\,.
\end{split}
\end{equation*}
Clearly $L_m \in \dt{2}$ and $L_{x,x} \in \dt{2}$; however, their intersection $L_{m,m}$ is not in $\dt{2}$, because its projection $\proj{L_{m,m}}{1} = \{ a^mb^m \}$ is not regular (i.e., in $\dt{1} = \ndt{1}$), contradicting the closure under projection established above.
Obviously, the proof generalizes to any $n \geq 2$.
Notice that $L_{x,x}$ is synchronous, hence $\dt{2}$ is not closed under intersection even with synchronous languages.

Since $\dt{n}$ is closed under complement but not under intersection, it is also not closed under union by De Morgan's laws: $L_1 \cap L_2 = \comp{\comp{L_1} \cup \comp{L_2}}$.
However, if $L_1, L_2 \in \dt{n}$ then $L_1 \cup L_2 \in \ndt{n}$, because a nondeterministic automaton can recognize $L_1$ and $L_2$ in parallel.

Fischer and Rosenberg showed~\cite[Sec.~3]{FischerR68} that $\dt{n}$ is not closed under concatenation, Kleene closure, or reversal.
The proof uses the three simple $2$-word languages $E = \{\langle a,a\rangle, \langle b,b\rangle\}$, $G = \{\langle c,c\rangle\}$, $H = \{\langle a,\epsilon\rangle, \langle b,\epsilon\rangle\}$.
Clearly, 
$$E^*GH^* = \{\langle xcy, xc \rangle \mid x,y \in \{a,b\}^* \} \in \dt{2}\,;$$
because recognizing $E^*GH^*$ amounts to comparing the words on the two tapes up to the character $c$; however, the concatenation
$$E^*GH^* \cat E^*GH^* = \{\langle x_1cy_1x_2cy_2, x_1cx_2c \rangle \mid x_1,x_2,y_1,y_2 \in \{a,b\}^* \}
$$
is not in $\dt{2}$, intuitively because a deterministic automaton cannot ``guess'' when $y_2$ ends and $x_2$ begins (formally, a counting argument similar to the pumping lemma of regular languages shows that a deterministic automaton must misclassify some words).
This implies that $\dt{2}$, and hence $\dt{n}$, is not closed under concatenation.
A similar arguments that considers arbitrary concatenations of $E^*GH^*$ proves that $\dt{2}$ is not closed under Kleene closure, either.

To show non-closure under reversal, take the language $H^*E^* = \{xy,y \mid x,y \in \{a,b\}^*\}$, which is accepted by a nondeterministic automaton in $\ndt{2}$ that guesses when $x$ ends in $xy$.
A modification of the argument used for $E^*GH^* \cat E^*GH^* \not\in \dt{2}$ proves that $H^*E^*$ is not recognizable with a deterministic automaton.
However, $\rev{H^*E^*} = E^*H^*$ clearly is in $\dt{2}$; in all $\dt{2}$, and $\dt{n}$, is not closed under reversal.

\subsubsection{Nondeterministic One-Way Automata}
\label{sec:nondet-one-way-automata-closures}
Let us consider the closure properties of the class $\ndt{n} = \nfa{n,\infty,0}$, for $n \geq 2$.

$\ndt{n}$ is closed under projection, using the same construction discussed in Section~\ref{sec:det-one-way-automata-closures} for deterministic automata: given an $n$-tape automaton $A$, build an $n-1$-tape automaton $B$ that replicates $A$'s behavior on all tapes except the $k$-th (which is projected out), where it behaves nondeterministically (that is, it performs any computation $A$ may perform on the $k$-th tape).
Hence if $L \in \ndt{n}$ then $\proj{L}{k} \in \ndt{n-1}$.

$\ndt{n}$ is closed under union; Elgot and Mezei gave~\cite{ElgotMezei65} the first proof of this fact (using a different formalism, ultimately corresponding to one-way multi-tape automata), which generalizes the construction for 1-tape automata: a nondeterministic automaton checks all components of a finite union in parallel through nondeterministic choice, and accepts if and only if at least one parallel computation accepts.
Similarly, the classic constructions for 1-tape automata are applicable to $n$-tape automata to prove that $\ndt{n}$ is closed under concatenation, Kleene closure, and reversal.

Like $\dt{n}$, $\ndt{n}$ is not closed under intersection.
The proof is the very same as in the deterministic case: $L_m$ and $L_{x,x}$ (Section~\ref{sec:det-one-way-automata-closures}) are in $\ndt{2}$ (even in $\dt{2}$), but their intersection $L_{m,m}$ is not in $\ndt{2}$ because its projection on either component is not regular, whereas $\ndt{n}$ is closed under projection.

Closure under union and non-closure under intersection imply, through De Morgan's laws, that $\ndt{n}$ is not closed under complement.

Finally, $\ndt{n}$ is not closed under generalization.
The proof uses the following lemma proved by Monks~\cite[Th.~4]{Monks10}: for every $n$-word language $L$ in $\ndt{n}$ there exists an $(n+1)$-word language $L'$ in $\dt{n}$ such that $L = \proj{L'}{1}$.
The proof gives an effective construction that turns an automaton accepting $L$ into a deterministic automaton accepting $L'$ over an extended alphabet, which can be projected out to obtain the original set $L$ of $n$-words.
Now, consider $L_m$ and $L_{x,x}$ again; both are in $\dt{2}$ and, since $\dt{2}$ is closed under complement, their complements $\comp{L_m}$ and $\comp{L_{x,x}}$ also are in $\dt{2}$, hence in $\ndt{2}$ \emph{a fortiori}.
$\ndt{2}$ is closed under union, thus the language
$$
L_{\cup} \ = \ \comp{L_m} \cup \comp{L_{x,x}}
$$
is also in $\ndt{2}$.
Therefore, the lemma we just stated ensures the existence of another language $\widetilde{L_{\cup}}$ in $\dt{3}$ such that $\proj{\widetilde{L_{\cup}}}{1}$ equals $L_{\cup}$; since $\widetilde{L_{\cup}} \in \dt{3}$, its complement $\comp{\widetilde{L_{\cup}}}$ is in $\dt{3}$ as well.
Now, notice that the complement $\comp{L_{\cup}}$ of $L_{\cup}$ is equivalent to
$$
\comp{L_{\cup}}\ =\ 
\comp{\comp{L_m} \cup \comp{L_{x,x}}} \ = \ 
L_m \cap L_{x,x} \ = \ 
L_{m,m} \,,
$$
which we noted is not in $\ndt{2}$.
If $\ndt{n}$ were closed under generalization,
$$
\fall{\comp{\widetilde{L_{\cup}}}}{1}
\ = \ 
\comp{\proj{\widetilde{L_{\cup}}}{1}}
\ = \ 
\comp{L_{\cup}}
$$
would be in $\ndt{2}$, which we noted is not the case.
In all, $\ndt{n}$ is not closed under generalization.

\subsubsection{Deterministic Rewind-Bounded Automata}
Let us consider the closure properties of the class $\rdt{n} = \dfa{n,\infty,\rb{\infty}}$, for $n \geq 2$.

$\rdt{n}$ is closed under complement, since each pass of a rewind automaton on the input tapes corresponds to a computation of an automaton in $\dt{n}$, which is closed under complement.
Hence, it is sufficient to adapt the usual construction for complementation in the following way: after every pass, accept if the original automaton rejects, reject if the original automaton accepts, and rewind if the original automaton rewinds.

$\rdt{n}$ is closed under union: given an automaton $A_1 \in \rdt{n}$ accepting $L_1$ and another automaton $A_2 \in \rdt{n}$ accepting $L_2$, an automaton $A$ accepting $L_1 \cup L_2$ operates as follows: simulate $A_1$; if $A_1$ accepts $A$ also accepts; otherwise, $A$ rewinds, simulates $A_2$, and accepts iff $A_2$ does.
Notice that termination is not a problem because the final part of Section \ref{sec:expr-incr-with-rewinds} discussed why we can assume that the number of rewinds is always bounded.

Closure under union and complement imply closure under intersection for $\rdt{n}$ through De Morgan's laws.
Alternatively, we can produce a direct construction similar to the one for union, where the simulation of two automata $A_1, A_2$ accepts iff both simulations accept.

Not only is $\rdt{n}$ closed under all Boolean operations; Rosenberg proves~\cite[Th.~8]{Rosenberg67} that every language in $\rdt{n}$ is expressible as a finite Boolean combination of languages in $\dt{n}$ (essentially, each of finitely many rewinds represents a one-way deterministic computation over the input), hence $\rdt{n}$ is the \emph{Boolean closure} of $\dt{n}$.

$\rdt{n}$ is not closed under concatenation, because we have already observed in Section~\ref{sec:nond-not-repl-rewinds} that the languages $B$ and $C$ are in $\dt{2}$ (hence in $\rdt{2}$ \emph{a fortiori}), but their concatenation $A = B \cat C$ is not in $\rdt{2}$ because accepting it requires nondeterminism. 
(As usual, the result immediately carries over to $\rdt{n}$).

Since, however, $\rev{A} = \{\langle b^max, b^m \rangle \mid m \in \naturals, x \in \{a,b\}^*\}$ clearly is in $\rdt{2}$, we conclude that $\rdt{n}$ is not closed under reversal either.

$\rdt{n}$ is not closed under Kleene closure; Rosenberg's proof~\cite[Th.~8]{Rosenberg67} works as follows.
Consider the language
$$
D \ =\ \left\{ \langle b^mc, b^m \rangle \mid m \in \naturals \right\}\,;
$$
clearly $D \in \rdt{2}$ (actually, even $D \in \dt{2}$).
Then, the same argument showing $B \cat C$ not in $\rdt{2}$ proves that $B \cat D \not \in \rdt{2}$.
Assume by contradiction that $\rdt{2}$ is closed under Kleene closure; then, since $\rdt{2}$ is closed under union and intersection, the language
\begin{equation}
A' \ =\ (B \cup D)^* \cap \big(B \cat (\{a,b\}^*)^n \cat \{\langle c, \epsilon \rangle \} \big)
\label{eq:A-prime}
\end{equation}
also is in $\rdt{2}$.
This is a contradiction, because $A'$ equals $B \cat D$---as the $B$ in the right-most argument of $\cap$ in \eqref{eq:A-prime} forces $B$ to appear first, and the closing $\langle c, \epsilon \rangle$ forces $D$ to appear second and last---but we know that $B \cat D \not\in \rdt{2}$.

We can prove that $\rdt{n}$ is not closed under projection using the language $L_{m,m} = L_m \cap L_{x,x}$ defined in Section~\ref{sec:det-one-way-automata-closures}.
We know that both $L_m$ and $L_{x,x}$ are in $\dt{2}$, hence their intersection $L_{m,m}$ is in $\rdt{2}$ (it is easy to build an automaton in $\rdt{2}$ that accepts $L_{m,m}$ with one rewind).
The projection $\proj{L_{m,m}}{1} = \{a^m,b^m\}$ is, however, not in $\rdt{1}$ because it is not regular and $\rdt{1} = \ndt{1}$.

Finally, $\rdt{n}$ is not closed under generalization, because it is closed under complement but not under projection.

\subsubsection{Nondeterministic Rewind-Bounded Automata}
Let us consider the closure properties of the class $\rndt{n} = \nfa{n,\infty,\rb{\infty}}$, for $n \geq 2$.

Using rewinds, the same constructions that show $\rdt{n}$ closed under intersection and union are applicable to show $\rndt{n}$ is closed under intersection and union: use one rewind to execute the two computations for $L_1$ and $L_2$ in parallel; accept if both accept, when recognizing $L_1 \cap L_2$; accept if at least one accepts, when recognizing $L_1 \cup L_2$.

$\rndt{n}$ is not closed under projection, with the same proof we have used in several other cases: $L_{m,m}$ is in $\rndt{2}$, but $\proj{L_{m,m}}{1} = \{a^mb^m\}$ is not in $\rndt{1}$ because it is not regular and $\rndt{1} = \ndt{1}$.

$\rndt{n}$ is not closed under complement either.
The proof is a modification of Chan~\cite[Th.~4.6]{Chan-PhD80}: consider the language
$$
L_{\neg\star,m} \ =\ 
\left\{
\left\langle\begin{array}{l}
u_1\star v_1 \# \cdots \# u_m\star v_m ,\\
x_1\star y_1 \# \cdots \# x_n\star y_n
\end{array}\right\rangle
\left|
\begin{array}{c}
\left(
\begin{array}{l}
u_1, \ldots, u_m, \\ 
v_1, \ldots, v_m, \\
x_1, \ldots, x_n, \\
y_1, \ldots, y_n 
\end{array}
\right)
\in \{0,1\}^+\,, \\
\forall i,j: u_i = x_j \;\Rightarrow\; v_i \neq y_j
\end{array}
\right.
\right\}\,.
$$
$L_{\neg\star,m}$ essentially is the complement of $L_{\star,m}$, but we can show that $L_{\neg\star,m} \not\in\rndt{2}$ whereas $L_{\star,m} \in \rndt{2}$.
Intuitively, rewinds do not help in recognizing $L_{\star,m}$ or $L_{\neg\star,m}$, as $m$ and $n$ can be arbitrarily large, whereas a finite-state automaton can only ``remember'' a finite number of position to jump to after any rewind; nontdeterminism works to recognize $L_{\star,m}$ but is still insufficient for $L_{\neg\star,m}$ where ``universal'' nondeterminism is needed.
Formally, assume by contradiction that $L_{\neg\star,m} \in\rndt{2}$ for some automaton $A$ with $r$ rewinds.
Then, modify $A$ into $A'$ so that $A'$ works on inputs of the form
$$
\langle x_1 \# x_2 \# \cdots \# x_{r+2}, y_1 \# y_2 \# \cdots \# y_{r+2} \rangle
$$
as $A$ would work on inputs of the form
$$
\langle a \star x_1 \# a^2 \star x_2 \# \cdots \# a^{r+2} \star x_{r+2}, 
a^{r+2} \star y_1 \# a^{r+1} \star y_2 \# \cdots \# a \star y_{r+2} \rangle\,,
$$
by having enough states to count $a$'s up to $r+2$.
Therefore $A'$ recognizes $L_{0/1,m}$ with $r$ rewinds, contradicting the fact that $L_{0/1,m}$ is not in $\nfa{2,\infty,r}$.
Since $m$ is generic, the contradiction shows that $L_{\neg\star,m} \not\in\rndt{2}$, then $\rndt{n}$ is not closed under complement.

$\rndt{n}$ is closed under concatenation, Kleene closure, and reversal, through the usual constructions that exploit nondeterminism mentioned in Section~\ref{sec:nondet-one-way-automata-closures}.

Finally, whether $\rndt{n}$ is closed under generalization is an open problem.

\subsubsection{Reversal-Bounded Automata}
The proof that establishes the non-closure under projection of automata with rewinds (by reduction to the 1-tape case) entails the non-closure under projection of automata with generic reversals in $\nfa{n,\infty,\ift{r}}$ and $\nfa{n,\infty,\ift{r}}$, for any $n \geq 2$ and $\ift{r} > 0$.
Since non-closure under projection of a class of automata is tantamount to undecidability of emptiness for automata of that class, we do not proceed further with the investigation of the closure properties of two-way automata (which have not been studied in the literature either).

\subsection{Decidability}
This section discusses the decidability of various problems for multi-tape automata.

\subsubsection{Deterministic One-Way Automata}\label{sec:dec-determ-one-way}
Let us consider a generic member $A$ of the class $\dt{n} = \dfa{n,\infty,0}$, for $n \geq 2$.

Emptiness is decidable, as a corollary of the fact that the projection of $A \in \dt{n}$ is in $\ndt{n-1}$, which is closed under projection.
Thus, $\lang{A}$ is empty if and only if $\proj{\proj{\cdots \proj{\lang{A}}{{t_{n-1}}}}{{t_2}}}{{t_1}}$ is empty, where $t_1, \ldots, t_{n-1}$ is any subset of $\{1, \ldots, n\}$ with $n-1$ elements; the latter problem is decidable because emptiness of regular languages is decidable.
Notice that the projection automaton is effectively constructible from $A$.
Another proof of the same decidability results uses an analogue of the pumping lemma for multi-tape automata: Rosenberg~\cite{Rosenberg-PhD} shows that, given a partition $Q_1 \cup \cdots \cup Q_n = Q$ of $A$'s set $Q$ of states, $\lang{A} \neq \emptyset$ if and only if $A$ accepts some input $\langle x_1, \ldots, x_n \rangle$ such that $|x_i| \leq |Q_i|$ for all $1 \leq i \leq n$.
Since $\dt{n}$ is closed under complement, universality is also decidable for $A \in \dt{n}$.

The property of projection also shows that finiteness is decidable for $A \in \dt{n}$: $\lang{A}$ is finite if and only if $\proj{\proj{\cdots \proj{\lang{A}}{{t_{n-1}}}}{{t_2}}}{{t_1}}$ is finite for \emph{every} subset $t_1, \ldots, t_{n-1}$ of $\{1, \ldots, n\}$ with $n-1$ elements; the latter problem is decidable because finiteness of regular languages is decidable.

The problem of whether equivalence is decidable for $A \in \dt{n}$ has been open for several years; Harju and Karhum\"aki~\cite{TJ-equivalence-dtn} have finally shown it to be decidable.
Their proof is quite technical, and it is basically a corollary of the related problem of ``multiplicity equivalence'' for nondeterministic one-way multi-tape automata.
Two automata in $\ndt{n}$ are \emph{multiplicity equivalent} if they accept the same $n$-words exactly the same number of times (that is with the same number of distinct computations).
Using some group-theoretic techniques, Harju and Karhum\"aki show that the multiplicity equivalence problem is decidable for automata in $\ndt{n}$, hence also for automata in $\dt{n}$.
The multiplicity equivalence problem reduces, however, to (ordinary) equivalence for deterministic automata in $\dt{n}$, which have at most one computation on each input.

Rabin and Scott~\cite[Th.~18]{RabinScott59} have shown that disjointness is undecidable for automata in $\dt{2}$; as usual, the proof immediately generalizes to $\dt{n}$ with $n \geq 2$.
The proof is a simple reduction from Post's correspondence problem: given a finite set of 2-words over $\Sigma$ (with $|\Sigma| \geq 2$)
$$
\left\{
\langle x_1, y_1\rangle,
\langle x_2, y_2\rangle,
\ldots,
\langle x_m, y_m\rangle
\right\}\,,
$$
determine if there exists a sequence $i_1, i_2, \ldots, i_k$ of indices from $1, \ldots, m$ (possibly with repetitions) such that
$$
x_{i_1} \,x_{i_2}\, \cdots\, x_{i_k} 
\ =\ 
y_{i_1} \,y_{i_2}\, \cdots\, y_{i_k}\,.
$$
The reduction is as follows: consider the (finite) 2-word languages
\begin{align*}
X \ &=\ \{ \langle x_1 , 1\rangle, \ldots, \langle x_m , m \rangle \}\,, \\
Y \ &=\ \{ \langle y_1 , 1\rangle, \ldots, \langle y_m , m \rangle \}\,.
\end{align*}
Obviously, $X^*$ and $Y^*$ are both in $\dt{2}$, but the given instance of Post's correspondence problem has a solution if and only if $X^* \cap Y^* \neq \emptyset$.
Since Post's correspondence problem is undecidable, disjointness for $\dt{2}$ (and $\dt{n}$) is also undecidable.

The undecidability of the inclusion problem for $\dt{n}$ follows from the fact that $\dt{n}$ is closed under complement but disjointness is undecidable for it; in fact, $A \subseteq B = \emptyset$ if and only if $A \subseteq \comp{B}$.

\subsubsection{Nondeterministic One-Way Automata}
Let us consider a generic member $A$ of the class $\ndt{n} = \nfa{n,\infty,0}$, for $n \geq 2$.

The proof of decidability of emptiness and finiteness for $\dt{n}$ does not depend on the restriction on deterministic automata, and in fact establishes that the same problems are decidable for $\ndt{n}$ as well.

Conversely, undecidability results for $\dt{n}$ immediately extend to its superclass $\ndt{n}$; thus disjointness and inclusion are undecidable for $\ndt{n}$ as well.

Universality is undecidable for $\ndt{n}$; Fischer and Rosenberg~\cite[Th.~7]{FischerR68} give another proof using reduction from Post's correspondence problem (introduced in Section~\ref{sec:dec-determ-one-way}: $X^* \cap Y^* = \emptyset$ if and only if $\comp{\comp{X^*} \cup \comp{Y^*}} = \emptyset$ if and only if $\comp{X^*} \cup \comp{Y^*} = (\Sigma^*)^2$, which is the universality problem for a language in $\ndt{2}$ (remember that $\ndt{n}$ is closed under union).

Now, notice that there exist an automaton $A_U$ in $\dt{n}$ (hence also in $\ndt{n}$) that accepts the universe language $(\Sigma^*)^n$.
Since deciding the equivalence $\lang{A} = \lang{A_U}$ is equivalent to deciding the universality problem for a generic $A \in \ndt{n}$, it follows that equivalence is also undecidable for $\ndt{n}$.

Finally, deciding whether a generic language $L \in \ndt{n}$ is deterministically recognizable (i.e., $L \in \dt{n}$) is also undecidable, with a more complex reduction from Post's correspondence problem~\cite[Th.~9]{FischerR68}.

\subsubsection{Deterministic Rewind-Bounded Automata}
All decision problems considered in this paper are undecidable for the class $\rdt{n} = \dfa{n,\infty,\rb{\infty}}$, for $n \geq 2$.
This is a consequence of the non-decidability of some problems for the class $\dt{n}$, of which $\rdt{n}$ is a strict superset, combined with the closure of $\rdt{n}$ under all Boolean operations.

Consider, for example, the emptiness problem for $\rdt{n}$; if it were decidable, then the disjointness problem for generic $A, B \in \dt{n}$ would be decidable by reducing it to testing the emptiness of $A \cap B \in \rdt{n}$.
But Section~\ref{sec:dec-determ-one-way} showed that disjointness is undecidable for $\dt{n}$.
Since $\rdt{n}$ is closed under complement, universality is also undecidable for $\rdt{n}$.

Emptiness is reducible to finiteness as follows. 
Assume, \emph{a contrario} that finiteness is decidable for $\rdt{n}$; then, we can decide whether $\lang{A} = \emptyset$ for a generic $A \in \rdt{n}$. 
First, determine if $A$ accepts a finite or infinite language; if it accepts an infinite language, we conclude that $\lang{A}$ is not empty; thus, let $\lang{A}$ be finite.
Modify $A$ into $A'$ by adding a loop on every accepting state of $A$; the loop reads some $\sigma \in \Sigma$ and moves the head right on every tape.
It should be clear that, for every $n$-word $\langle x_1, \ldots, x_n \rangle$ accepted by $A$, $A'$ accepts all (infinitely many) words in 
$$
\left\{ 
\langle x_1 \sigma^m, x_2 \sigma^m, \ldots, x_n \sigma^m \rangle
\mid
m \in \naturals
\right\}\,.
$$
We have that $A'$ accepts a finite language if and only if $\lang{A}$ is empty, but emptiness is undecidable for a generic $A \in \rdt{n}$.

Disjointness is undecidable for $\rdt{n}$ because the universe language $(\Sigma^*)^n$ is in $\rdt{n}$, and deciding disjointness of $\lang{A}$ and $(\Sigma^*)^n$ is equivalent to deciding emptiness of $\lang{A}$.
Similarly, $\lang{A} \subseteq \emptyset$ if and only if $\lang{A} = \emptyset$, hence inclusion is also undecidable for $\rdt{n}$.
Finally, equivalence is a generalization of emptiness, thus it is undecidable for $\rdt{n}$.

\subsubsection{Other Two-Way Automata}
The undecidability of all decision problems for $\rdt{n}$ propagates up the hierarchy: all decision problems are undecidable for $\rndt{n}$, as well as for generic two-way automata.

\section{Summary of Properties} \label{sec:summary-properties}

Figure~\ref{fig:expr} summarizes the known inclusion between classes of multi-tape automata. Solid lines denote strict inclusions and dotted lines denote weak inclusions; namely, it is unknown whether nondeterminism increases the expressive power of asynchronous multi-tape automata with unbounded reversals.

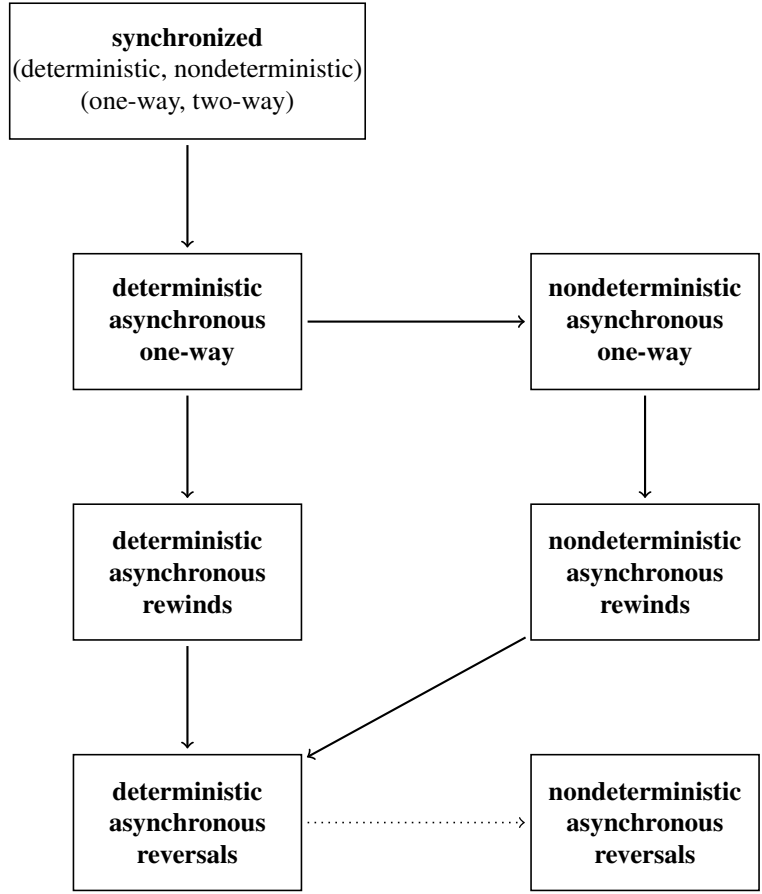
\begin{figure}[!hb]
\centering
\begin{tikzpicture}[->,semithick,draw,inner sep=1pt,
  shorten >=2pt, shorten <=2pt, node distance=15mm and 30mm,align=center,
  every node/.style={draw,shape=rectangle,minimum height=18mm,minimum width=30mm}]

\node (reg) {\textbf{synchronized} \\ (deterministic, nondeterministic) \\ (one-way, two-way)};
\node (dao) [below=of reg] {\textbf{deterministic} \\ \textbf{asynchronous} \\ \textbf{one-way}};
\node (nao) [right=of dao] {\textbf{nondeterministic} \\ \textbf{asynchronous} \\ \textbf{one-way}};

\node (dar) [below=of dao] {\textbf{deterministic} \\ \textbf{asynchronous} \\ \textbf{rewinds}};
\node (darw) [below=of dar] {\textbf{deterministic} \\ \textbf{asynchronous} \\ \textbf{reversals}};

\node (nar) [below=of nao] {\textbf{nondeterministic} \\ \textbf{asynchronous} \\ \textbf{rewinds}};
\node (narw) [below=of nar] {\textbf{nondeterministic} \\ \textbf{asynchronous} \\ \textbf{reversals}};

\begin{scope}[thick,->]
\path (reg) edge (dao);
\path (dao) edge (dar);
\path (dar) edge (darw);

\path (dao) edge (nao);
\path (nao) edge (nar);
\path (nar) edge (darw);
\end{scope}

\path (darw) edge [dotted] (narw);

\end{tikzpicture}
\caption{Relative expressiveness of multi-tape automata classes.}
\label{fig:expr}
\end{figure}

Table~\ref{tab:closure} summarizes the known closure properties of synchronized multi-tape automata (\textbf{syn}), deterministic asynchronous one-way (\textbf{da1}), nondeterministic asynchronous one-way (\textbf{na1}), deterministic asynchronous with reversals (\textbf{dar}), and non deterministic asynchronous with reversals (\textbf{nar}), with respect to complement $\neg$, intersection $\cap$, union $\cup$, projection $\exists$, generalization $\forall$, concatenation $\cat$, Kleene star $^*$, and reversal $\mathit{rev}$.
``Y'' means closure, ``N'' means non-closure, ``n'' denotes closure within the next class, and ``?'' means open problem.

\begin{table}[!h]
\centering
\begin{tabular}{cccccccc}
& \textbf{syn} &  \textbf{da1}  &  \textbf{na1}  &  \textbf{dar}  &  \textbf{nar} \\
\hline
$\neg$          & Y & Y & N & Y & N \\
$\cap$          & Y & N & N & Y & Y \\
$\cup$          & Y & N & Y & Y & Y \\
$\exists$       & Y & n & Y & N & N \\
$\forall$       & Y & N & N & N & ? \\
$\cat$          & Y & N & Y & N & Y \\
$^*$            & Y & N & Y & N & Y \\
$\mathit{rev}$  & Y & N & Y & N & Y 
\end{tabular}
\caption{Closure properties of multi-tape automata.}
\label{tab:closure}
\end{table}

Table~\ref{tab:decidability} shows which problems are decidable (``Y'') and which undecidable (``N'') for the same classes of multi-tape automata.

\begin{table}[!h]
\centering
\begin{tabular}{cccccccc}
& \textbf{syn} &  \textbf{da1}  &  \textbf{na1}  &  \textbf{dar}  &  \textbf{nar} \\
\hline
Emptiness          & Y & Y & Y & N & N \\
Universality       & Y & Y & N & N & N \\
Finiteness         & Y & Y & Y & N & N \\
Disjointness       & Y & N & N & N & N \\
Inclusion          & Y & N & N & N & N \\
Equivalence        & Y & Y & N & N & N
\end{tabular}
\caption{Decidability properties of multi-tape automata.}
\label{tab:decidability}
\end{table}

\clearpage
\section{Intersection of Asynchronous Automata} \label{sec:intersection}
This section describes an algorithm for the intersection of multi-tape nondeterministic asynchronous one-way finite-state automata.
Since these are not closed under intersection, the algorithm may not terminate (or, equivalently, it may define an infinite-state automaton as result).
We use a slightly different definition of multi-tape automaton, which is easily seen equivalent to Definition~\ref{def:NDA} in the one-way case.

\begin{definition} \label{def:n-tape-aut}
An \emph{$n$-tape finite-state automaton} $A$ is a tuple $\langle \Sigma, T, Q, \tau, \delta, Q_0, F\rangle$ where: $\Sigma$ is the input alphabet, with $\emk \not\in \Sigma$; $T = \{t_1, \ldots, t_n\}$ is the set of tapes; $Q$ is the finite set of states; $\tau: Q \to T$ assigns a tape to each state; $\delta: Q \times \Sigmaend \to \powerset{Q}$ is the (nondeterministic) transition function; $Q_0 \subseteq Q$ are the initial states; $F \subseteq Q$ are the accepting (final) states.
\end{definition}
Whenever convenient we will represent the transition function $\delta$ as a relation, that is the set of triples $(q, \sigma, q')$ such that $q' \in \delta(q, \sigma)$.

Consider two asynchronous automata $A = \langle \Sigma, Q^A, \delta^A, Q^A_0, F^A, T^A, \tau^A \rangle$ and $B = \langle \Sigma, Q^B, \delta^B, Q^B_0, F^B, T^B, \tau^B \rangle$, such that $A$ has $m$ tapes $T^A = \{t_1^A, \ldots, t_m^A\}$ and $B$ has $n$ tapes $T^B = \{t_1^B, \ldots, t_n^B\}$.
We now describe an algorithm that computes the intersection $C = A \cap B$ of $A$ and $B$, where $C = \langle \Sigma, Q, \delta, Q_0, F, T, \tau \rangle$; $C$'s tapes $T$ are the union of $T^A$ and $T^B$.
To describe the algorithm, we introduce repeated operations as separate routines.
All components of the algorithm have access to the definitions of $A$ and $B$ and to a global stack \lstinline|s| where new states of the composition are pushed (when created) and popped (when processed).

Routine \lstinline|async_next| takes a $t$-tape automaton $D$ (i.e., $A$ or $B$) and one of its states $q$, and returns a set of tuples $\langle q', h_1, \ldots, h_t \rangle$ of all next states reachable from $q$ either directly or by accumulating delayed transitions $h_i \in (\delta^D)^*$ in tape $t_i$, for $1 \leq i \leq t$.
We call \emph{delayed states} such tuples of states with delayed transitions.
The search for states reachable from $q$ stops at the first occurrences of states associated with a certain tape.
Figure~\ref{fig:asyncNext} shows the pseudo-code for \lstinline|async_next|.

\begin{figure}[!h]
\begin{lstlisting}
async_next (D, q): SET [$\langle q', h_1, \ldots, h_t \rangle$]
   -- $q$ is always reachable from itself
   Result $:= \{ \langle q, \epsilon, \ldots, \epsilon\rangle \}$
   -- for every tape other than $q$'s
   for each $t_i \in \{t_1^D, \ldots, t_t^D\} \setminus \tau^D(q)$ do
      $P := \text{all shortest paths } p \text{ from }q \text{ to some }\overline{q}  \text{ such that: }$
            $\tau^D(\overline{q}) = t_i \text{ and no state } \widetilde{q} \text{ with } \tau^D(\widetilde{q}) = t_i \text{ appears in } p \text{ before } \overline{q}$
      -- each element in $P$ is a sequence of transitions
      for each $e_1\,\cdots\,e_m \in P$ do
         $h_1, \ldots, h_t := \epsilon$
         -- each transition is a triple (source, input, target)
         for each $(q_1, \sigma, q_2) \in e_1\,\cdots\,e_m$ do
            -- add the transition to the sequence corresponding
            -- to its source's tape
            $h_{\tau^D(q_1)} := h_{\tau^D(q_1)} + (q_1, \sigma, q_2)$
         -- $q_2(e_m)$ is the target state of the last transition $e_m$
         Result $:=$ Result $\cup \langle q_2(e_m), h_1, \ldots, h_t\rangle$
\end{lstlisting}
\caption{Routine \lstinline|async_next|.}
\label{fig:asyncNext}
\end{figure}

Consider now a pair of delayed states $\langle p, h_1, \ldots, h_m \rangle$ and $\langle q, k_1, \ldots, k_n \rangle$, respectively for automata $A$ and $B$.
The two delayed states can be composed only if the delays on the synchronized tapes are pairwise \emph{consistent}, that is the sequence of input symbols of one is a prefix (proper or not) of the other's; \lstinline|cons($h_i, k_i$)| denotes that the sequences $h_i, k_i$ of delayed transitions are consistent.
Routine \lstinline|new_states| (in Figure~\ref{fig:newStates}) takes two sets $P, Q$ of delayed states and returns all consistent states obtained by composing them.
\lstinline|new_states| also pushes onto the stack \lstinline|s| all composite states that have not already been added to the composition.
For convenience, \lstinline|new_state| also embeds the tape $t$ of each new composite state within the state itself.

\begin{figure}[!h]
\begin{lstlisting}
new_states (P: SET[$\langle p, h_1, \ldots, h_m \rangle$], Q: SET[$\langle q, k_1, \ldots, k_n \rangle$]): S
   S $:= \emptyset$
   for each $\langle p, h_1, \ldots, h_m \rangle \in P$, $\langle q, k_1, \ldots, k_n \rangle \in Q$ do
      -- if delays on synchronized tapes are consistent
      if $\forall i \in T^A \cap T^B:$ cons($h_i, k_i$) then
         for each $t \in T$ do $S := S \cup \{\langle p, q, t, h_1, \ldots, h_m, k_1, \ldots, k_n \rangle \}$ end
   -- Here $Q$ denotes $C$'s set of states, not the input argument
   for each $r \in S$ do if $r \not\in Q$ then s.push (r) end
\end{lstlisting}
\caption{Routine \lstinline|new_states|.}
\label{fig:newStates}
\end{figure}

It is often convenient to add arbitrary prefixes to the delays of delayed states generated by \lstinline|new_states|. To this end, routine \lstinline|compose_transition| (in Figure~\ref{fig:composeTransition}) takes two sets $P, Q$ of delayed states and an $(m+n)$-tuple of delays, and calls \lstinline|new_states| on the modified states obtained by orderly adding the delays to the states in $P$ and $Q$.
It also adds all transitions to the newly generated states to the transition function $\delta$ of the composite $C$.

\begin{figure}[!h]
\begin{lstlisting}
compose_transition (P: SET[$(p, h_1, \ldots, h_m)$], Q: SET[$(q, k_1, \ldots, k_n)$], 
                    d: ($h_1, \ldots, h_m, k_1, \ldots, k_n$), $\sigma$, r)
   $J_A := \{ (p, h_1\,h_1', \ldots, h_m\,h_m') \mid (p, h_1', \ldots, h_m') \in P \}$
   $J_B := \{ (q, k_1\,k_1', \ldots, k_n\,k_n') \mid (q, k_1', \ldots, k_n') \in Q \}$
   $S :=$ new_states ($J_A, J_B$)
   for each $r' \in S$ do $\delta := \delta \cup \{r, \sigma, r'\}$ end
\end{lstlisting}
\caption{Routine \lstinline|compose_transition|.}
\label{fig:composeTransition}
\end{figure}

We are ready to show the main routine \lstinline|intersect| which builds $C$ from $A$ and $B$.
Since the intersection may have infinite states, \lstinline|intersect| takes as arguments a bound on the maximum number of states and on the maximum delay (measured in number of transitions) accumulated in the states.
\clearpage
\begin{lstlisting}
intersect (max_states, max_delay)
  $Q := \emptyset$ ; $s := \emptyset$
  -- $A$'s initially reachable states
  $J_A := \bigcup_{i \in I_A}$ async_next (A, i)
  -- $B$'s initially reachable states
  $J_B := \bigcup_{i \in I_B}$ async_next (B, i)
  S $:=$ new_states ($J_A$, $J_B$)
  -- mark these states as initial
  $I := S$
  until $s = \emptyset$ or $|Q| \geq$ max_states loop
    r $:=$ ($q_a, q_b, t, h_1, \ldots, h_m, k_1, \ldots, k_n$) $=$ s.pop
    if $\forall d \in \{ h_1, \ldots, k_n \}: |d| \leq $ max_delay then 
      $Q := Q \cup \{ r \}$ 
    else continue
    $Q := Q \cup \{ r \}$
    if $t \in T^A \cap T^B$ then
      -- event on shared tape
      if $h_t = (u_a, \sigma, u_a') \overline{h_t}$ and $k_t = (u_b, \sigma, u_b') \overline{k_t}$ then
        -- delayed transition on both $A$ and $B$
        $P :=$ async_next $(A, q_a)$
        $Q :=$ async_next $(B, q_b)$
        $d := (h_1, \ldots, \overline{h_t}, \ldots, h_m, k_1, \ldots, \overline{k_t}, \ldots, k_n)$
        compose_transition ($P, Q, d, \sigma, r$)
      elseif $h_t = (u_a, \sigma, u_a') \overline{h_t}$ and $k_t = \epsilon$ then
        -- delayed transition on $A$
        $P :=$ async_next $(A, q_a)$
        -- normal transition on $B$
        $Q := \{$ async_next ($B, q_b'$) $\mid (q_b, \sigma_b, q_b') \in \delta^B \land \sigma = \sigma_b \land \tau^B(q_b) = t \}$
        $d := (h_1, \ldots, \overline{h_t}, \ldots, h_m, k_1, \ldots, k_n)$
        compose_transition ($P, Q, d, \sigma, r$)
      elseif $h_t = \epsilon$ and $k_t = (u_b, \sigma, u_b') \overline{k_t}$ then
        -- delayed transition on $B$
        $Q :=$ async_next $(B, q_b)$
        -- normal transition on $A$
        $P := \{$ async_next ($A, q_a'$) $\mid (q_a, \sigma_a, q_a') \in \delta^A \land \sigma = \sigma_a \land \tau^A(q_a) = t \}$
        $d := (h_1, \ldots, h_m, k_1, \ldots, \overline{k_t}, \ldots, k_n)$
        compose_transition ($P, Q, d, \sigma, r$)
      elseif $h_t = k_t = \epsilon$ then
        for each $\sigma \in \Sigma$ do
          -- normal transition on both $A$ and $B$
          $P := \{$ async_next ($A, q_a'$) $\mid (q_a, \sigma_a, q_a') \in \delta^A \land \sigma_a = \sigma \land \tau^A(q_a) = t \}$
          $Q := \{$ async_next ($B, q_b'$) $\mid (q_b, \sigma_b, q_b') \in \delta^B \land \sigma_b = \sigma \land \tau^B(q_b) = t \}$
          $d := (h_1, \ldots, h_m, k_1, \ldots, \ldots, k_n)$
          compose_transition ($P, Q, d, \sigma, r$)
    elseif $t \in T^A \setminus T^B$ then
      -- event on $A$'s non-shared tape
      if $h_t = (u_a, \sigma, u_a') \overline{h_t}$ then
        -- delayed transition on $A$, $B$ stays
        $P :=$ async_next $(A, q_a)$
        $Q := \{ (q_b, \epsilon, \ldots, \epsilon) \}$
        $d := (h_1, \ldots, \overline{h_t}, \ldots, h_m, k_1, \ldots, k_n)$
        compose_transition ($P, Q, d, \sigma, r$)
      elseif $h_t = \epsilon$ then
        -- normal transition on $A$, $B$ stays
          $Q := \{ (q_b, \epsilon, \ldots, \epsilon) \}$
          for each $\sigma \in \Sigma$ do
            $P := \{$ async_next ($A, q_a'$) $\mid (q_a, \sigma_a, q_a') \in \delta^A \land \sigma_a = \sigma \land \tau^A(q_a) = t \}$
            $d := (h_1, \ldots, h_m, k_1, \ldots, k_n)$
            compose_transition ($P, Q, d, \sigma, r$)
    elseif $t \in T^B \setminus T^A$ then
      -- event on $B$'s non-shared tape
      if $k_t = (u_b, \sigma, u_b') \overline{k_t}$ then
        -- delayed transition on $B$, $A$ stays
        $P := \{ (q_a, \epsilon, \ldots, \epsilon) \}$
        $Q :=$ async_next $(B, q_b)$
        $d := (h_1, \ldots, h_m, k_1, \ldots, \overline{k_t}, \ldots, k_n)$
        compose_transition ($P, Q, d, \sigma, r$)
      elseif $k_t = \epsilon$ then
        -- normal transition on $B$, $A$ stays
          $P := \{ (q_a, \epsilon, \ldots, \epsilon) \}$
          for each $\sigma \in \Sigma$ do
            $Q := \{$ async_next ($B, q_b'$) $\mid (q_b, \sigma_b, q_b') \in \delta^B \land \sigma_b = \sigma \land \tau^B(q_b) = t \}$
            $d := (h_1, \ldots, h_m, k_1, \ldots, k_n)$
            compose_transition ($P, Q, d, \sigma, r$)
\end{lstlisting}

\clearpage

\end{document}